\documentclass{aa}  

\usepackage{graphicx}
\usepackage{txfonts}
\usepackage{amsmath}
\usepackage{graphicx}
\usepackage{dcolumn}
\usepackage{bm}
\usepackage{url}
\usepackage[caption=false]{subfig}
\usepackage{tikz} 

\def\apj{Astrophys. J.}
\def\apjl{Astrophys. J.\ letts.}
\def\apjs{Astrophys. J. \ Suppl.}

\def\grl{Geophys. Res. \ Lett.}

\let \aap = \aa

\usepackage{listings}
\usepackage{color}
\definecolor{dkgreen}{rgb}{0,0.6,0}
\definecolor{gray}{rgb}{0.5,0.5,0.5}
\definecolor{mauve}{rgb}{0.58,0,0.82}

\begin{document}

   \title{A comparison of methods for finding magnetic nulls in simulations and in situ observations of space plasmas}


    \author{V. Olshevsky\inst{1}\fnmsep\inst{2}
            \and
            D. I. Pontin\inst{3}\fnmsep\thanks{Corresponding author}
            \and
            B. Williams\inst{4}
            \and
            C. E. Parnell\inst{4}
            \and
            H. S. Fu\inst{5}
            \and
            Y. Liu\inst{5}
            \and
            S. Yao\inst{6}\fnmsep\inst{7}
            \and
            Y. V. Khotyaintsev\inst{8}
    }

    \institute{KTH Royal Institute of Technology, 
               Lindstedtsvagen 5, SE-10044 Stockholm, Sweden\\
        \and
               Main Astronomical Observatory of NAS, 
               Akademika Zabolotnoho 27, 03680, Kyiv, Ukraine
        \and
               School of Mathematical and Physical Sciences,
               University of Newcastle, University Drive, Callaghan,
               NSW 2308, Australia\\
               \email{David.Pontin@newcastle.edu.au}
        \and
               University of St Andrews, North Haugh, St Andrews, UK\\
        \and
               Beihang University,
               37 Xueyuan Road,
               Haidian District, Beijing 100191, China\\
        \and
               Institute of Space Sciences,
               Shandong University,
               Weihai, China\\
        \and
               State Key Laboratory of Space Weather,
               National Space Science Center,
               Chinese Academy of Sciences,
               Beijing, China
        \and
               Swedish Institute of Space Physics,
               Box 537, SE-751 21 Uppsala, Sweden\\
    }

   \date{Received September ??, 2020; accepted ????}

 
  \abstract
   {Magnetic nulls are ubiquitous in space plasmas, and are of interest as sites of localised energy dissipation or magnetic reconnection. As such, a number of methods have been proposed for detecting nulls in both simulation data and \emph{in situ} spacecraft data from Earth's magnetosphere. The same methods can be applied to detect stagnation points in flow fields.}
   {In this paper we describe a systematic comparison of different methods for finding magnetic nulls. The Poincar\'e index method, the first-order Taylor expansion (FOTE) method, and the trilinear method are considered.}
   {We define a magnetic field containing fourteen magnetic nulls whose positions and types are known to arbitrary precision. Furthermore, we applied the selected techniques in order to find and classify those nulls. Two situations are considered: one in which the magnetic field is discretised on a rectangular grid, and the second in which the magnetic field is discretised along synthetic `spacecraft trajectories' within the domain.}
   {At present, FOTE and trilinear are the most reliable  methods for finding nulls in the spacecraft data and in numerical simulations on Cartesian grids, respectively. The Poincar\'e index method is suitable for simulations on both tetrahedral and hexahedral meshes.}
   {The proposed magnetic field configuration can be used for grading and benchmarking the new and existing tools for finding magnetic nulls and flow stagnation points.}

            
   \titlerunning{Methods for finding magnetic nulls}

   \maketitle
%

\section{Introduction}

   Astrophysical plasmas are typically characterised by high magnetic Reynolds numbers, and their magnetic fields are found to exhibit a complex structure on a range of scales. For example, observations from missions studying the Earth's magnetosphere (Cluster \citep{Escoubet:2001} and the Magnetospheric Multiscale (MMS) mission \citep{Burch:2016}) show highly fluctuating fields both in the magnetotail \citep{Fu:etal:2016} and magnetosheath \citep{Retino:etal:2007}. Extrapolations of the solar coronal magnetic field based on photospheric magnetograms similarly show enormous complexity in the magnetic connectivity between photospheric flux fragments \citep[e.g.][]{schrijver2002,close2003}. In order to understand the detailed dynamics of such highly complex fields, we need to identify the features of the magnetic field at which localised energy conversion -- typically mediated by magnetic reconnection -- takes place. Candidates for locations of magnetic reconnection in complex 3D fields include magnetic nulls (points at which the magnetic field strength, ${\bf B}$, is zero), their associated separatrix surfaces, and the separator lines that are formed by the intersections of these separatrix surfaces \citep[for a review, see e.g.][]{pontin2012,priest2014}. Such magnetic nulls have been detected in spacecraft data from the magnetotail, magnetopause, magnetosheath, and foreshock {\citep[e.g.][]{Xiao:etal:2006NatPh,He:2008,deng2009,wendel2013,Guo:2016a,Fu:etal:2018,Chen:etal:2017,Chen:etal:2019}}. In extrapolations of the solar coronal field they are found in abundance, with the number of nulls increasing exponentially as the photosphere is approached \citep[][]{longcope2009}. Moreover, magnetic reconnection at these nulls has been implicated in energy release in, for example, solar flares and jets \citep[e.g.][]{masson2009,yang2015,kumar2019}. Being an isolated point, a magnetic null is not easy to detect in discrete data. This has motivated the development of methods which infer the existence of magnetic null points, both in simulations and in spacecraft data. 

   Methods for finding topological singularities and other special features are becoming increasingly important for researchers working with huge amounts of simulation and observational data. A topological analysis is extremely useful for observers as specific features (e.g.~magnetic null points) are likely locations for energetic events in the Sun or the Earth's magnetosphere (for appropriate external perturbations). The same analysis allows us to distinguish the important subsets of the huge amounts of data collected from satellites or telescopes. A topological analysis serves two main purposes when applied to numerical simulations: It allows the identification and classifications of the data sets (or simulation sub-domains) of potential interest, and it could also be used for data compression. Finally, as discussed above, in both observations and simulations, certain topological features have very important physical implications and serve as a framework to understand the physical processes that drive the observed dynamics.

  This paper addresses a subclass of topological analysis techniques, namely the identification of the stagnation points of 3D vector fields. In the present discussion (around space and astrophysical plasmas), we mainly discuss the topology of the magnetic field. The same analysis, however, is applicable to other vector fields, such as the flow velocity \citep{Wang:etal:2020}, so long as the field is divergence-free. We designed and undertook a  `challenge' to compare the performance of different null finding approaches to understand possible deficiencies and weaknesses of several pieces of software. Limited comparisons were  previously made between different methods \citep{Fu:etal:2020}, but they either did not include as many methods, or they did not include a `ground truth' in which the exact existence and positions of the nulls are known \citep{haynesparnell2007, Eriksson:etal:2015}. Our aim is to understand how the choice of method could possibly influence the analysis of observations or simulations, and what are each method's strengths and weaknesses. We compare the most popular methods in the literature that can be automated to quickly analyse many different instances of input data (see Section \ref{sec:methods}).


\section{Theoretical background}
\label{sec:background}
\subsection{Field structure in the vicinity of a magnetic null}
  Magnetic nulls are locations in space at which the magnetic field is zero, and in the generic case this occurs at isolated points. The structure of the magnetic field in the vicinity of these points can be characterised by linearising the field about the point. We note that for any generic (stable) null this linearisation is non-zero, and the local topology of the field linearisation can be shown to be the same as the local topology of the field itself -- see \cite{Hornig:1996}. The eigenvalues and eigenvectors of the magnetic field Jacobian $\nabla{\bf B}$ at the null determine the \emph{spine-fan} structure of the field, as described in detail in \cite{Fukao:1975,Parnell:1996}. Since $\nabla\cdot{\bf B}=0$, the eigenvalues sum to zero. The eigenvectors associated with the two same-sign eigenvalues locally define a plane in which magnetic field lines approach or recede from the null, known as the \emph{fan} surface (or $\Sigma$-surface). The remaining eigenvector defines the direction of the \emph{spine} line (or $\gamma$-line), along which field lines recede from or approach the null. If the same-sign eigenvalues have negative real parts, the null has topological degree $+1$ -- in the literature this is sometimes termed either an \emph{A-type} null or \emph{negative} null. If the same-sign eigenvalues have positive real parts, the topological degree is $-1$, and we have a \emph{B-type} null or \emph{positive} null. One further pertinent distinction is between nulls for which all three eigenvalues are real (\emph{radial nulls}), and those for which two eigenvalues are complex conjugates (\emph{spiral nulls}). In the latter case, the field lines form a spiral pattern in the fan surface, and nulls are sometimes denoted as being of $A_s$- or $B_s$-type (this occurs when a sufficiently strong component of electric current is present parallel to the spine line).

  \subsection{Test field for the challenge}
\begin{figure}
\begin{center}
	\includegraphics[width=0.9\columnwidth]{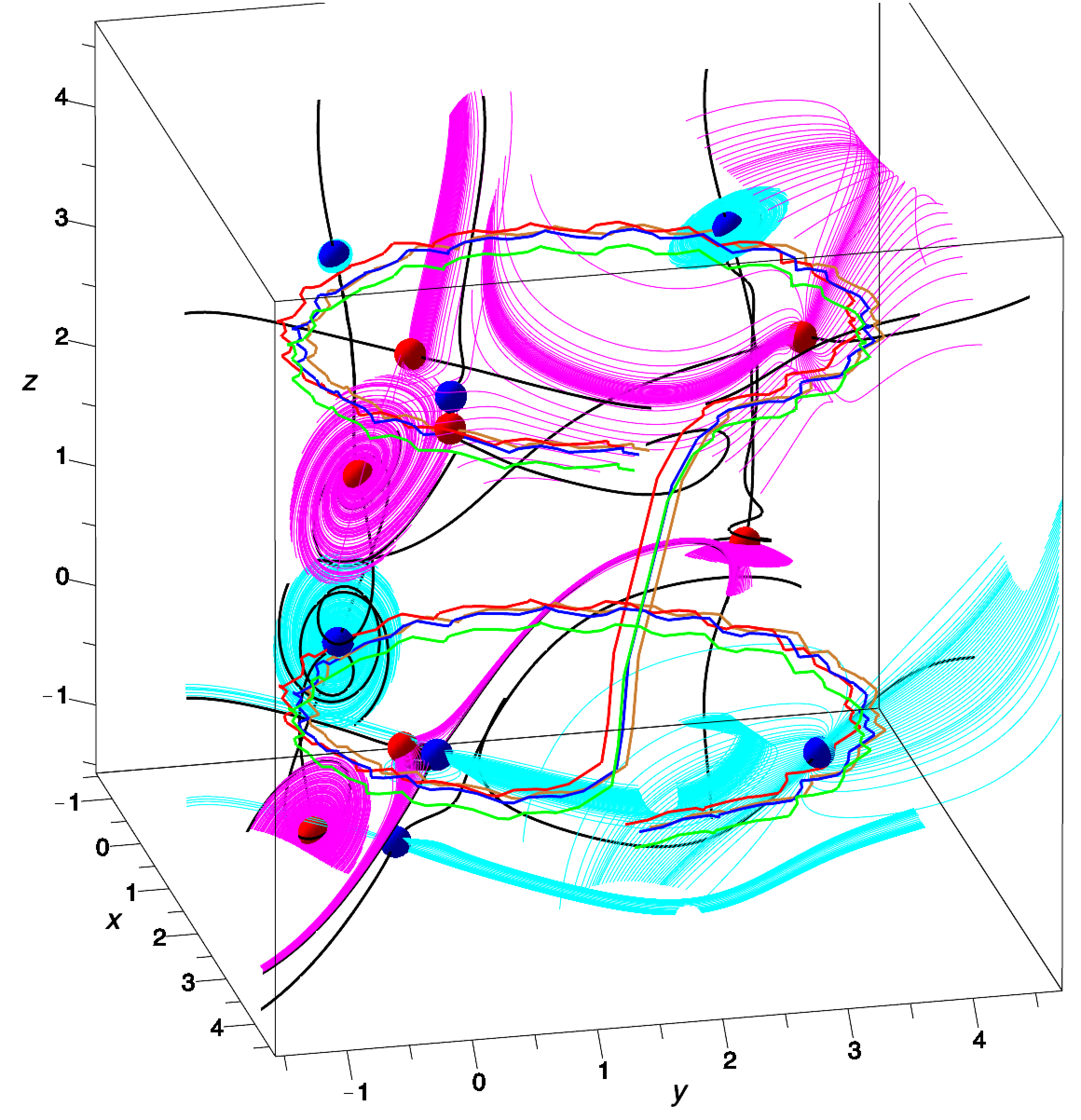}
	\caption{Visualisation of the test magnetic field showing the null points and associated field line structures, together with the simulated spacecraft trajectories. Null points of topological degree $+1$ ($-1$) lie at the centre of {blue (red)} spheres and their fan field lines are represented in {cyan (magenta)}. Spine lines are black. The red, blue, green and gold curves are the simulated spacecraft trajectories.
	}
	\label{fig:nulls_and_trajectories}
\end{center}
\end{figure}

  The magnetic configuration used to test the null finding methods is based on a triply-periodic field that has previously been used to initiate turbulence simulations \citep{Politano:1995}. To this field various perturbations are added in order to make the disposition of null points less `regular'. Some of these perturbations take the form of `flux rings', that are inserted in such a way as to lead to a pitchfork bifurcation of one of the pre-existing nulls, leading to the creation of two additional nulls \citep[as in e.g.][]{Wyper:2014a}.  This is done in such a way that all nulls can be accounted for based on theoretical arguments. Following the addition of the perturbations as described, the exact null locations can no longer be obtained analytically. Instead they are obtained using Newton's method. Since the field itself is still known analytically, the null location can still be found to arbitrary precision. Further, since the field is prescribed analytically, the Jacobian of ${\bf B}$ can also be calculated exactly at these points, and thus the topological degree of the null and the local orientation of the spine line and fan surface can be determined as above. Details for constructing the magnetic field are presented in the Appendix~\ref{app:setup}. The null points and their spine and fan structures are represented within the volume of interest ($x,y,z\in [-\pi/2,3\pi/2]$) in Figure \ref{fig:nulls_and_trajectories}.

  \subsection{Data sets used to test methods}\label{sec:dataset}
  The different null finding methods are described in the following section. These are designed to be used to find nulls on either hexa- or tetrahedral meshes of data points (obtained from numerical simulations), or on time-series of quadruplets of measurements (taken by the Cluster or MMS spacecraft). We therefore generate two different types of data sets from the model magnetic field. In the first case, we evaluate the magnetic field components on a rectilinear grid of points with various different resolutions. In the second, we define four trajectories through the domain, evaluating the magnetic field components at a discrete set of points along these trajectories (see Fig.~\ref{fig:nulls_and_trajectories}). These points are chosen in each instance to lie at the corners of a regular tetrahedron, to mimic typical spacecraft configurations. The trajectories are designed so as to pass close to some of the null points, and further from others -- see the Appendix~\ref{app:setup} for details. We compare the results for {three} different sizes of tetrahedron (corresponding to different spacecraft separations). 
  
  The magnetic field as defined in {Appendix \ref{app:setup}} is dimensionless. {Particularly in the context of the trajectory data from hypothetical spacecraft tetrahedra, it is relevant to compare with physical length scales. One possible dimensionalisation would be to consider our domain to be equivalent to the largest fully-kinetic simulations afforded by present-day codes and supercomputers, since these were recent data sets on which null finders were applied. Those simulations consider domains extending for tens of ion inertial lengths $d_i$ \citep{Pucci:etal:2017ApJ,Olshevsky:etal:2018ApJ}. If we suppose that the domain size is $20\,d_i$ in each dimension, then our `small-scale' and `medium-scale' tetrahedra have spacecraft separations of $0.13d_i$ and $0.63d_i$, respectively. The inter-spacecraft separation of the MMS constellation can change between $5-80km$ which corresponds to $0.005-0.08d_i$ in the magnetotail, and $0.05-0.8d_i$ in the magnetopause. Hence, our `small-scale' tetrahedron aims at resembling the electron diffusion region scales covered by the MMS mission. The inter-spacecraft separation of the Cluster mission varies $200-2500km$, resembling $0.2-2.5d_i$ in the magnetotail, and $2-25d_i$ in the magnetopause. This larger separation -- of the order of the ion diffusion region -- dictates the choice of the `medium' and `large' scales for our study.}
  


  \section{Methods}\label{sec:methods}
  This section describes the three methods that we have compared, their theoretical formulation and implementation.

  \subsection{Poincar\'e index method}
  \subsubsection{Theoretical formulation}
  The problem of locating a magnetic null is essentially a problem of finding a root of a continuous divergence-free vector field. The Poincar\'e index or topological degree method for finding such roots was introduced by \cite{Greene:1992}. This technique has been applied to various kinetic simulations by \cite{Olshevsky:etal:2015ApJ,Olshevsky:2016ApJ} and spacecraft observations by { \citet{Eriksson:etal:2015,Xiao:etal:2006NatPh}.} The key assumption of the method consists in the linearity of the field around a null, therefore a field in the neighbourhood of the null is given by 
\begin{equation}
B_i = \left(\mathbf{\nabla B}\right)_{ij}\left(x_j - x_{j0}\right),
\end{equation} 
  where the summation is implied over repeating indices, $x_{j0}$ denote the coordinates of the null, and $\left(\mathbf{\nabla B}\right)=\partial B_i / \partial x_j |_{x_j=x_j0}$ is the magnetic field gradient, a $3\times 3$ matrix constant. The topological degree of the field in the specified volume of space is represented by the following sum over all nulls
\begin{equation}\label{eq:index}
D=\sum_{k}\textrm{sign}\left[ \lambda_{1k} \lambda_{2k} \lambda_{3k} \right],
\end{equation}
  where $\lambda_{1k,2k,3k}$ are the eigenvalues of the $\mathbf{\nabla B}$ at the $k$-th null. As mentioned above, in the generic case nulls do not degenerate (in reality they can be degenerate only at one instant of time during a bifurcation process), and all three eigenvalues are non-zero. The implication of this fact is that magnetic nulls are isolated. As the topological degree is strongly conserved, it provides a measure of the difference between the number of positive and the number of negative nulls in the given volume of space. If the volume of space is sufficiently small, one can assume it encloses exactly one null if $D \ne 0$.

  The topological degree over a region of space can be evaluated from the field on the surface which encloses this region of space. Typically the magnetic field is given at the vertices of a cell, either hexahedral or tetrahedral. Each face of the cell is split into triangles (see Fig.~8c in \citet{Fu:etal:2015JGR}), and the field in the centre of each triangle is interpolated from its corners. In this way, we translate from the `configuration space' into the `magnetic field space'. To find out if the cell's surface encloses a null of the magnetic field, each triangular face is mapped onto a unit sphere in the magnetic field space. The area of each triangle's projection on the unit sphere is given by the solid angle subtended by the three magnetic field vectors. The sum of the areas of all triangles ($4$ for tetrahedron and $12$ for hexahedron), divided by $4\pi$, gives the number of times the triangles cover a unit sphere in the magnetic field space. This is the sum of the signs of all the nulls of the field inside the sphere (see Eq.~(\ref{eq:index})). We note that  each area has a sign, and it is important to observe the order of vertices in the triple cross product $\mathbf{B}_1\cdot \mathbf{B}_2 \times \mathbf{B}_3$ to get the sign correctly. The `plus' sign corresponds to the outward directed flux, while the `minus' refers to the inward field flux.

  For our implementation of the Poincar\'e index method we use the formula for the solid angle subtended by three vectors proposed by \cite{VanOosterom:Strackee:1983IEEE}:
\begin{eqnarray}
\tan\left(\frac{1}{2}\Omega\right)= 
\frac{\mathbf{B}_1\cdot \mathbf{B}_2 \times \mathbf{B}_3}{B_1 B_2 B_3 + \left( \mathbf{B}_1\cdot\mathbf{B}_2 \right) B_3   + \left( \mathbf{B}_1\cdot\mathbf{B}_3 \right) B_2  + \left( \mathbf{B}_2\cdot\mathbf{B}_3 \right) B_2}.
\end{eqnarray}
  Evaluation of the solid angle this way is faster and more stable than the conventional use of the Cosine theorem. In particular, there is no need for zero-denominator checks when using modern programming languages, as errors are handled by the \lstinline|arctan2| function.

  Once a cell which encloses a null is found, it is possible to get a more precise estimate of the null location inside this cell using the Secant theorem \citep{Greene:1992}. However, as noted by \citet{Greene:1992}, this estimate may often be misleading, even giving locations outside the cell. Our experiments confirmed this problem, therefore it is more practical to assume the null is located in the centre of mass of the cell.
  
  The topological classification of a null is straightforward on hexahedral cells where finite differences can be used to deduce the magnetic field Jacobian. A technique for $\mathbf{\nabla B}$ computation in tetrahedral cells is given in \cite{Khurana:etal:1996IEEE}.

  \subsubsection{Implementation}
  The null-finder based on the Poincar\'e index method\footnote{https://bitbucket.org/volshevsky/magneticnullchallenge} combines magnetic field measurements into a set of either $4$ or $8$ magnetic field vectors given in the vertices of a cell. It computes the topological degree and returns either a very small number close to zero (meaning no null is present inside the cell), or a number close to $1$ or $-1$, meaning there is a null inside the cell. In practice, the thresholds of being `close to zero' or `close to one' are regulated by the numerical precision. Similarly to the trilinear method described in Section \ref{sec:tri_method}, only those grid cells are selected for analysis, in which none of the components of the magnetic field have the same sign at all vertices (this being incompatible with the existence of a null in the cell). If at least one component of the magnetic field has the same sign in all the vertices, the field can't go to zero inside this cell (in the linear approximation). This pre-selection reduces the computation cost dramatically in a typical simulation or observation, where only a fraction of measurements comprise field nulls.

  \subsection{Trilinear method}
  \label{sec:tri_method}
  \subsubsection{Theoretical formulation}
  The trilinear method for finding the locations of null points in a numerical grid under the trilinear assumption was originally formulated by \cite{haynesparnell2007}. The algorithm described below differs from \cite{haynesparnell2007} by using a deca-section method (like the bisection method) rather than Newton-Raphson method for converging on the location of the null points. There are three stages to the method: the reduction stage, bilinear stage and the sub-grid stage.

\begin{description}
    \item The reduction stage just reduces the amount of work done by the algorithm in the bilinear stage. Each grid cell is checked in turn for  a change in sign in the magnetic field. Essentially, a grid cell cannot contain a null point under the trilinear assumption if all 8 values of the grid cell vertices are of the same sign (see above).
    \item The bilinear stage actually checks for the possibility of a null point within a grid cell. The zero isosurfaces ($ B_i = 0 $) of the three components of the magnetic field will intersect at a null point if a null exists. This triple intersection is difficult to find directly numerically. However, two of the three components of the magnetic field's zero isosurfaces will intersect to form a line which the null point must lie on. This line must also intersect with the grid cell faces. The magnetic field on these grid cell faces is now only bilinear and therefore the locations of the intersection points of this line and the cell faces (say $P_1$ and $P_2$) can be found analytically. The values of the third component of the magnetic field (unused to form the line) can be found at $P_1$ and $P_2$: if a null point exists, then this third magnetic field component must be of opposite signs at $P_1$ and $P_2$. By using this test, the algorithm can detect which grid cells may contain null points.
    \item The final, sub-grid stage is simply the first two stages which are repeated at sub-grid cell resolution to identify the locations of the null points at the required accuracy. Each null-containing grid cell is split into a new grid of $ 10 \times 10 \times 10 $ cells where the magnetic field values are found using the trilinear assumption and the reduction and bilinear methods are applied to these smaller grid cells. This process of splitting each cell is repeatedly applied until the desired accuracy in the location is obtained.
\end{description}

  The methods used for detecting the sign of the null points use a convergence-style method. They are fully detailed in \cite{williams2018}. A field line about a null point can be written as
    \begin{equation}
        \label{eq:nullfline} \vec{r} \!\left(s\right) = a_1 e^{\lambda_1 s} \vec{e}_1 + a_2 e^{\lambda_2 s} \vec{e}_2 + a_3 e^{\lambda_3 s} \vec{e}_3
    \end{equation}
  where $ \lambda_i $ and $ \vec{e}_i $ are the corresponding eigenvalues and eigenvectors of $ M = \boldsymbol{\nabla} \vec{B} $ evaluated at the null point and $ a_i $ are constants.
  By repeated multiplication of equation (\ref{eq:nullfline}) by $ M $ (and assuming that $ \lambda_1 $ is the eigenvalue corresponding to the eigenvector associated with the spine line), we obtain
    \begin{equation}
        M^n \cdot \vec{r} \! \left(s\right) \to a_1 \lambda_1^n e^{\lambda_1 s} \vec{e}_1.
    \end{equation}
  This allows the eigenvector associated with the spine line to be identified. This convergence is used by the Sign Finder to classify the signs of the null points. This also identifies the eigenvectors associated with the fan plane. However, the Sign Finder does not find any of the values of eigenvalues of the null point or identify if it is a spiral null point. If this information is desired, it must be found by an alternative method.

  \subsubsection{Implementation}
  The algorithm for finding null points using the trilinear method is implemented in the Magnetic Skeleton Analysis Tools. It is a Fortran based package for analysing the skeleton of divergence-free vector fields\footnote{https://github.com/benmatwil/msat}.

  \subsection{FOTE method}
  \subsubsection{Theoretical formulation}
  The first-order Taylor expansion (FOTE) method is based on Taylor expansion of the magnetic field in the vicinity of a null \citep{Fu:etal:2015JGR}:
\begin{equation}
\textbf{B}\left(\textbf{r} \right) = \nabla \textbf{B} \left( \textbf{r} - \textbf{r}_0 \right),
\label{eq:FOTE-main}
\end{equation}
  where $\nabla \textbf{B}$ is the Jacobian matrix derived from four-point measurements, $\textbf{r}_0$ is the location of one of the four spacecraft, $\textbf{r}$ is the location of interest, and $\textbf{B} \left( \textbf{r} \right)$ is the magnetic field at the location of interest. Requiring $\textbf{B} \left( \textbf{r} \right)=0$ and inverting this linear expansion (Eq.~\ref{eq:FOTE-main}), we can obtain the null position r. In general, the equation will always give a position of a magnetic null, if the four spacecraft do not measure exactly the same magnetic field, which is impossible in observations or simulations where instrumental or numerical noise is inevitable. However, we only regard the null as reliably identified if (1) the null-spacecraft distance $\left( \textbf{r} - \textbf{r}_0 \right) < d_i$, (2) the following dimensionless error parameters are both smaller than $0.4$: 
\begin{eqnarray}
\eta = \left| \frac{ \nabla \cdot \textbf{B}}{\max( \nabla \textbf{B})}\right|, \\
\zeta = \left| \frac{ \lambda_{1} + \lambda_{2} + \lambda_{3} }{ \max(| real(\lambda) | )} \right|,
\end{eqnarray}
  where where $\lambda_1$, $\lambda_2$, and $\lambda_3$ are the eigenvalues of the Jacobian matrix $\nabla \textbf{B}$. The quantitative criteria for qualifying FOTE are derived from the comprehensive tests of the simulation data \citep{Fu:etal:2015JGR}.

  \subsubsection{Implementation}
  The algorithm for finding null points and identifying their type using the FOTE method is implemented in Matlab. The time-series quadruple data are used in such test. At each sampling point, a null point position relative to the spacecraft is calculated by Equation~\ref{eq:FOTE-main}. Since the spacecraft trajectories generated artificially are given, a null point location in the spatial domain can be obtained. 

  As we have introduced, owing to the linear assumption, the identification of a remote null point is not reliable. Thus, we set a threshold distance. Only the magnetic nulls below such threshold distance are further evaluated.
 
  \subsubsection{FOTE and SOTE methods}
  FOTE method has shown great powers in automatic null detection. However, the FOTE method requires the magnetic fields around the spacecraft to be quasi-linear, so that its accuracy is reduced when dealing with strongly non-linear magnetic fields. 

  Recently, a new method entitled `Second-Order Taylor Expansion' (SOTE) was proposed by \cite{Liu:2019:SOTE} to overcome the linear limitation. This method is based on the second-order Taylor expansion of the magnetic field
\begin{equation}
\mathbf{B}\left(x,y,z,t\right)=\mathbf{a}x + \mathbf{b}y + \mathbf{c}z + \mathbf{d}xy + \mathbf{e}xz + \mathbf{f}yz + \mathbf{l}x^2 + \mathbf{m}y^2 + \mathbf{n}z^2+\mathbf{B}_0,
\label{eq:SOTE}
\end{equation}
  where $\mathbf{a}$, $\mathbf{b}$, $\mathbf{c}$, $\mathbf{d}$, $\mathbf{e}$, $\mathbf{f}$, $\mathbf{l}$, $\mathbf{m}$, $\mathbf{n}$, and $\mathbf{B}_0$ are vector coefficients. The following constraints can be applied to these equations: $\mathbf{\nabla}\cdot \mathbf{B}=0$ and $\mathbf{\nabla}\times\mathbf{B}=u \mathbf{J}$, where the current density is derived from particle moments: $\mathbf{J}=ne\left(\mathbf{V}_i - \mathbf{V}_e\right)$. To completely determine all the coefficients in Equation~\ref{eq:SOTE}, the SOTE method utilises two sets of four-point measurements of magnetic field and particles, by assuming the structures to be quasi-stationary and solving the precise trajectory of the spacecraft.

  The SOTE method is good at reconstructing  non-linear structures, for example, the null-point pairs in this study. Thus, it enables the analysis of null-point pairs in space plasmas. However, for null point detection, the SOTE method should have the same performance as FOTE method since the FOTE reconstruction is essentially a local approximation of SOTE reconstruction. What's more, the SOTE method cannot be applied to a time-varying structure while the FOTE method could reveal the temporal evolution of a magnetic structure.


  \section{Results}
  The methods considered here can be broken down into two categories: Some methods are designed to take eight data points as input (meaning 24 data values when the three component of ${\bf B}$ are included), motivated by the need to find magnetic nulls in simulation data utilising rectilinear grids of points. Both the Trilinear and Poincar\'e methods have been applied previously in this way. By contrast, other methods were developed to find nulls in data from the four-spacecraft missions, Cluster and MMS, and therefore take as input the magnetic field components measured at four points in space: the FOTE and Poincar\'e index methods were used in this context. In the following sections we consider these two cases separately.

  \subsection{Results for eight-point methods}
  \subsubsection{Results of the Poincar{\' e} index method}
\begin{figure}
\begin{center}
	\includegraphics[width=0.98\columnwidth]{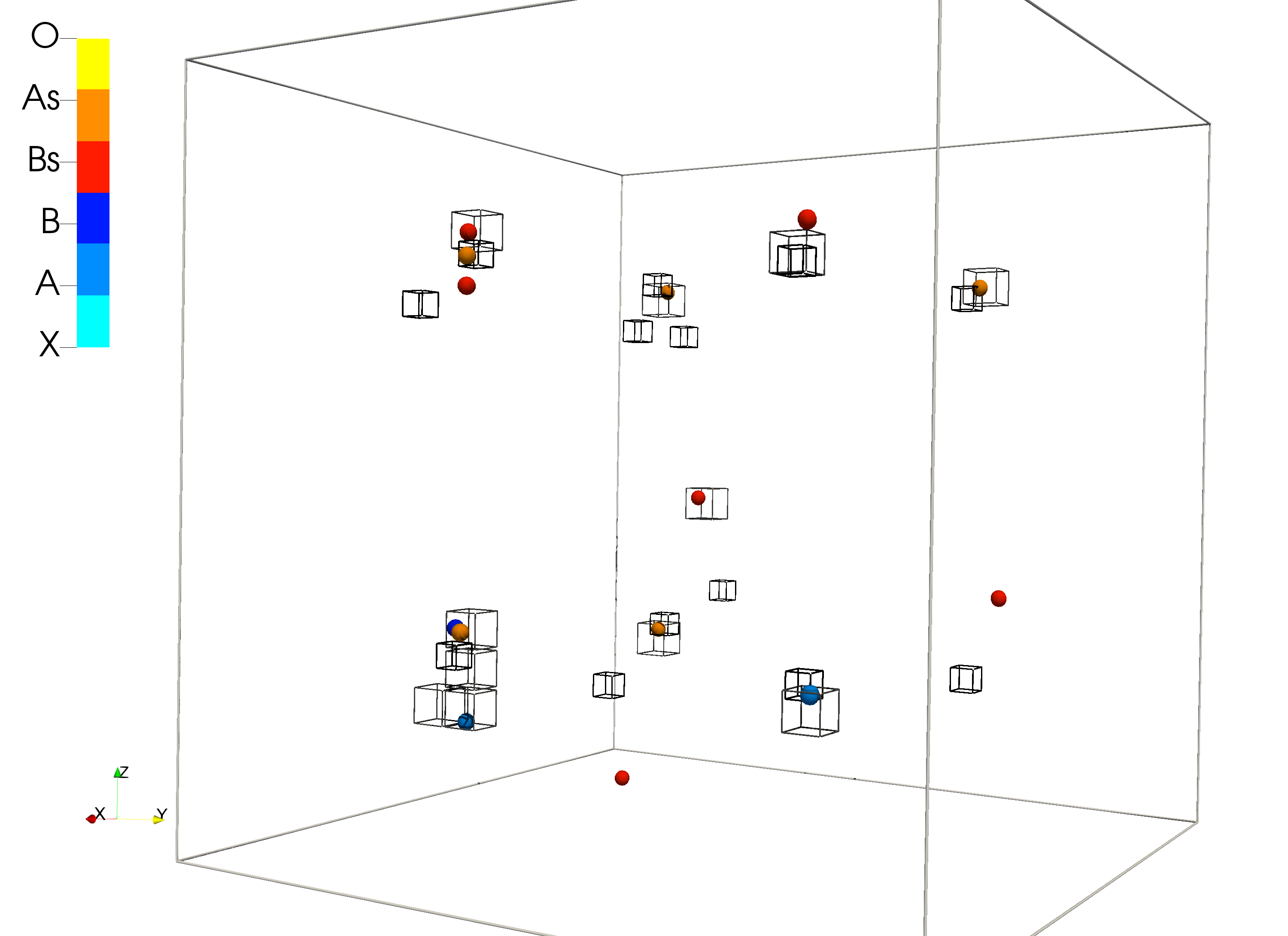}
	\caption{Locations of magnetic nulls found by the Poincar{\' e} index method for different grid resolutions. Cubic boxes outline the grid cells where nulls are detected in the $20^3$ grid (larger boxes), $30^3$ grid (smaller boxes). Colour spheres show the nulls found in the $80^3$ grid. Colour denotes the topological type of the null as described in the legend. }
	\label{fig:topology:poincare}
\end{center}
\end{figure}
  In Figure~\ref{fig:topology:poincare} we plot the locations of nulls found when the magnetic field is evaluated on equally-spaced grids with resolutions $20^3$, $30^3$, and $80^3$. We see that the detection and location of some of the nulls is relatively stable between the different resolutions, while other null detections exhibit large differences for the different resolutions. The null points detected at $20^3$ and $80^3$ resolution are listed in Table \ref{tab:tri_top_compare}. This is discussed further below.

  \subsubsection{Results of Trilinear method}

\begin{figure}
    \centering
    \includegraphics[width=0.98\columnwidth]{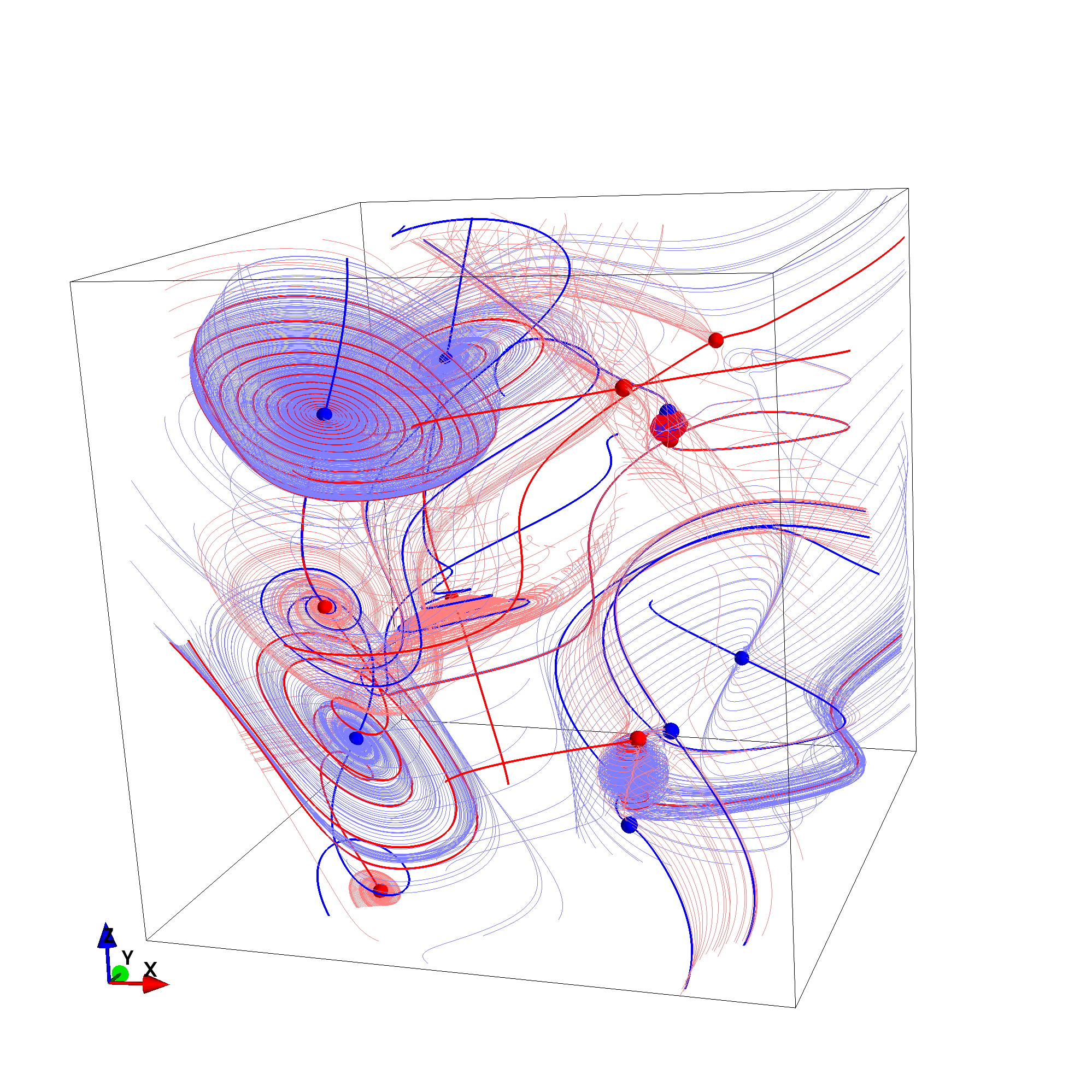}
    \caption{3D rendering of the magnetic field obtained using the trilinear method and associated Magnetic Skeleton Analysis Tools, {for $80^3$ resolution}. Positive and negative null points are represented as red and blue spheres respectively, spine lines from positive and negative null points are represented as thick red and blue lines respectively and the field lines originating from the fan planes of positive and negative null points are drawn as thinner red and blue lines respectively.}
    \label{fig:404040_3d_tri}
\end{figure}
\begin{figure}
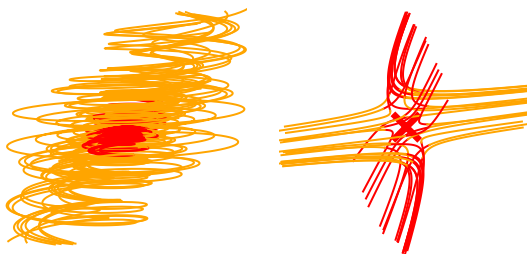

\centering
    \resizebox{0.4\columnwidth}{!}{\label{fig:badnull_tri}\input{badnull-tri1.pgf}}
    \resizebox{0.4\columnwidth}{!}{\label{fig:goodnull_tri}\input{goodnull-tri1.pgf}}
    \caption{Illustration of the effect of different grid resolution for the reconstructed magnetic field structure using the trilinear method. The magnetic field in the vicinity of the same null point is reconstructed from the $20^3$ grid data (left) and the $30^3$ grid data (right). The reconstructed field structure changes from a sink to a divergence-free null point when the resolution is increased. Red and orange field lines are traced in the forward and backward direction, respectively, of the magnetic field.}
\label{fig:nulls_tri}
\end{figure}

  Figure~\ref{fig:404040_3d_tri} illustrates a typical output of the magnetic skeleton analysis and application of the trilinear method to a $ 80^3 $ grid. Table \ref{tab:tri_top_compare} shows the results by using the trilinear method to find the locations of the null points in the test magnetic field on different resolution grids. In the $ 20^3 $ grid, the trilinear method only finds 12 null points and is only able to classify 11 of these. It cannot locate two of the 14 null points which exist in the analytical field.

  More closely analysing the unclassified null point reveals that, under the trilinear assumption, this point is represented by a sink with the field lines twisting into the point (Figure \ref{fig:nulls_tri}). However, it turns out that when the magnetic field is evaluated on a grid with $ 30^3 $ resolution, this point is revealed to be a true divergence-free null point (Figure \ref{fig:nulls_tri}). There is clearly just a resolution issue and the vector field is not approximately trilinear locally to this null point at $ 20^3 $ resolution.

  The two null points which cannot be located in the $ 20^3 $ grid are actually located within the same grid cell at $ 20^3 $ resolution. From analysis in the higher resolution grids (where these two null points are now located in different grid cells), we find one of these null points is positive and the other is negative. Since this pair of null points comprises both a positive and a negative null point, together they have a topological degree of zero and so they cannot be found when in the same grid cell.

  At $ 30^3 $ resolution, all 14 null points are now found using the trilinear method. However there is still one null which the algorithm is unable to classify. The exact same change as above occurs with null point 12 between $ 30^3 $ and $ 40^3 $ resolution. The field lines around null point 12 show that it appears to be a source in the trilinear approximation at $ 30^3 $ resolution and becomes a negative divergence free null point at $ 40^3 $ resolution. At $ 40^3 $ resolution and higher, all 14 null points are found and classified correctly.

  \subsubsection{Comparison between Poincar\'e index and trilinear methods}

\begin{table*}[!hbt]
\begin{center}
    \caption{Comparison of nulls locations and types obtained using the trilinear and Poincar\'e index methods to the true locations and types, for two different grid resolutions.}
    \label{tab:tri_top_compare}
    \begin{tabular}{c|cccc|cccc|cccc}
        \hline\hline
        {Null \#} & Type & $ x $ & $ y $ & $ z $  & Type & $ x $ & $ y $ & $ z $ & Type & $ x $ & $ y $ & $ z $ \\
        \hline
        $20^3$ & \multicolumn{4}{c}{Trilinear} &  \multicolumn{4}{c}{Poincar\'e} &  \multicolumn{4}{c}{Exact (2 d.p.)} \\
        \hline
        1  & B & 0.30 & -0.25 &-1.49   &   -  &   -   & -  &-  &$  B_s $&$ 0.30 $&$ -0.25 $&$ -1.48 $ \\ 
        2  & A & 0.04 & -0.00 &-0.01   &   $A_s$ & -0.08 & -0.08 & -0.08 &$ A_s$& $0.05 $&$ 0.00 $&$ -0.01 $  \\
        3  & A & 2.90 & -0.16 &-0.57   &   $A$  &  2.89 & -0.41 & -0.41 &$  A $&$2.87 $&$-0.19$&$-0.60$ \\ 
        4  & 0 & 2.95 & -0.17 & 0.17   &   $A_s$ &  2.89 & -0.08 & -0.08 &$  B_s $&$2.97$&$  -0.15$&$ 0.22$ \\ 
        5  & A & 3.29 &  0.10 & 0.40   &   $A_s$ &  2.89 & -0.08 &  0.25 &$  A_s $&$3.23 $&$  0.06$&$ 0.25 $ \\ 
        6  & B &-0.20 &  3.31 & 0.42   &    - &     -  &    -   &  -     &$  B_s $&$ -0.21 $&$ 3.32 $&$ 0.44 $  \\ 
        7  & B &-0.28 &  0.25 & 1.26   &   $B_s$ & -0.41 &  0.25 &  1.24  &$  B_s $&$ -0.27$&$ 0.24 $&$ 1.25 $ \\ 
        8  & A & 3.14 &  3.14 &-0.04   &   $A$  &  3.22 &  3.22 & -0.08 &$  A $&$3.14$&$  3.13$&$  -0.04 $ \\ 
        9  & A & 0.02 & -0.02 & 3.22   &   $A_s$ & -0.08 &  -0.08 &  3.22 &$  A_s $&$0.02 $&$ -0.02 $&$ 3.21 $ \\ 
        10&-&-&-&-&-&-&-&- &$  B_s $&$ 3.32$&$  0.15$&$ 2.99 $\\ 
        11 & A &-0.00 &  3.14 & 3.17   &   $A_s$ & -0.08 &  3.22 &  3.22 &$  A_s $&$ -0.00 $&$ 3.14$&$ 3.17 $ \\ 
        12&-&-&-&-&-&-&-&- &$  A_s $&$ 3.31$&$  0.15$&$ 3.25 $\\ 
        13 & B & 2.95 & -0.13 & 3.61   &   $B_s$ &  2.89 & -0.08 &  3.55  &$  B_s $&$2.99$&$  -0.10$&$  3.50 $ \\ 
        14 & B & 3.02 &  3.01 & 3.38   &   $B$  &  2.89 &  2.89 &  3.22  &$B_s$&$3.02$&$  3.02$&$  3.39 $ \\ 
        - & - &  -   &   -   &   -     &   $B_s$ &  2.89 & -0.08 & -0.41  &-&-&-&- \\
        \hline
        $80^3$ & \multicolumn{4}{c}{Trilinear} &  \multicolumn{4}{c}{Poincar\'e}  & \multicolumn{4}{c}{Exact (2 d.p.)} \\
        \hline
        1  & B & 0.30 & -0.25 &-1.48   &   $B_s$ &  0.30 & -0.26 & -1.45 &$  B_s $&$ 0.30 $&$ -0.25 $&$ -1.48 $ \\ 
        2  & A & 0.05 & -0.00 &-0.01   &   $A_s$ &  0.06 & -0.02 & -0.02 &$ A_s$& $0.05 $&$ 0.00 $&$ -0.01 $  \\
        3  & A & 2.86 & -0.19 &-0.60   &   $A$  &  2.84 & -0.18 & -0.58 &$  A $&$2.87 $&$-0.19$&$-0.60$ \\ 
        4  & B & 2.98 & -0.15 & 0.23   &   $B$  &  3.00 & -0.18 &  0.22 &$  B_s $&$2.97$&$  -0.15$&$ 0.22$ \\ 
        5  & A & 3.23 &  0.06 & 0.25   &   $A_s$ &  3.24 &  0.06 &  0.22 &$  A_s $&$3.23 $&$  0.06$&$ 0.25 $ \\ 
        6  & B &-0.21 &  3.32 & 0.44   &   $B_s$ & -0.18 &  3.32 &  0.46 &$  B_s $&$ -0.21 $&$ 3.32 $&$ 0.44 $  \\ 
        7  & B &-0.27 &  0.24 & 1.25   &   $B_s$ & -0.26 &  0.22 &  1.25 &$  B_s $&$ -0.27$&$ 0.24 $&$ 1.25 $ \\ 
        8  & A & 3.14 &  3.13 &-0.04   &   $A$  &  3.16 &  3.16 & -0.02 &$  A $&$3.14$&$  3.13$&$  -0.04 $ \\ 
        9 & A & 0.02 & -0.02 & 3.21   &   $A_s$ & -0.02 & -0.02 &  3.24 &$  A_s $&$0.02 $&$ -0.02 $&$ 3.21 $ \\ 
        10  & B & 3.32 &  0.15 & 3.00   &   $B_s$ &  3.32 &  0.14 &  3.00 &$  B_s $&$ 3.32$&$  0.15$&$ 2.99 $\\ 
        11 & A &-0.00 &  3.14 & 3.17   &   $A_s$ & -0.02 &  3.16 &  3.16 &$  A_s $&$ -0.00 $&$ 3.14$&$ 3.17 $ \\ 
        12 & A & 3.31 &  0.15 & 3.24   &   $A_s$ &  3.32 &  0.14 &  3.24 &$  A_s $&$ 3.31$&$  0.15$&$ 3.25 $\\ 
        13 & B & 2.99 & -0.10 & 3.51   &   $B_s$ &  3.00 & -0.10 &  3.48 &$  B_s $&$2.99$&$  -0.10$&$  3.50 $ \\ 
        14 & B & 3.02 &  3.02 & 3.39   &   $B_s$ &  3.00 &  3.00 &  3.40 &$B_s$&$3.02$&$  3.02$&$  3.39 $ \\ 
	\hline\hline
	\end{tabular}
\end{center}
\end{table*}

  The null point locations and types for the Poincar\'e index and trilinear methods are compared to the true values in Table \ref{tab:tri_top_compare}. Before comparing these results it is worth making some important notes. First,
 the trilinear method as currently implemented does not distinguish spiral nulls, since it does not make use of the Jacobian matrix eigenvalues to determine the null type (see above). In principle this could be done in the same way as for the Poincar\'e method (taking finite differences over the grid to evaluate the Jacobian matrix).
 Second, the Poincar\'e index method as currently implemented does not seek the nulls at sub-grid resolution, thus the centre of the cell is reported as the null location. By contrast, the trilinear method fits a field to the data, with the null point of this fitted field within the cell being reported.

  With the above in mind we can compare the nulls found by the two methods, in Table \ref{tab:tri_top_compare}. First, we see that at $20^3$ resolution, both methods are imperfect. The trilinear method misses two nulls (10 and 12 -- though note that as mentioned above these exist in the same grid cell at this resolution), while one fails the classification process. The Poincar\'e index method performs a little worse: in addition to the two nulls missed by the trilinear method, nulls 1 and 6 are also not found, while these is a false-positive null as well (bottom row of the table). The problematic null 4 is again wrongly classified, while the spiral nature of null 14 is not picked up.

  At $80^3$ resolution both methods show much better results, as expected. Both methods find all 14 nulls, with only null 4 wrongly classified by the Poincar\'e index method (again as $B$ rather than $B_s$). As a result of the fact that the field is fitted on the grid allowing some sub-grid resolution to the null detection, the trilinear method generally gives a more accurate estimate of the null point location.

  \subsection{Results for four-point methods}
  
  \begin{table}
\begin{center}
	\caption{FOTE method applied to measurement along trajectories. {The final column ($|R|$) gives the closest approach of the centre of the tetrahedron to the null.}}
	\label{tab:fote}
	\begin{tabular}{*6c}
    	\hline\hline
        {Index} & Type & x & y & z & {$|R|$} \\
        \hline
        \multicolumn{3}{l}{medium-scale: $S=0.12$} &  \multicolumn{3}{c}{Nulls found: $14$} \\
		\hline
		1 & X & 2.9372 & -0.1778 & 0.1631 & 0.1294\\
		2 & A & 3.1385 & 3.1349 & -0.0326 & 0.1282\\ 
		3 & B & 3.3098 & 3.5301 & 2.8745 & 0.6195\\
		4 & B & 3.3493 & 2.9556 & 2.9583 & 0.9480\\
		5 & O & -0.1163 & 3.1190 & 0.1774 & 0.5218\\
		6 & O & 0.0454 & -0.0002 & -0.0109 & 0.0276\\
		7 & O & 4.1302 & 0.8054 & 1.7989 & 1.7318\\
		8 & O & -0.0011 & 3.1369 & 3.1722 & 0.0726\\
		9 & O & 0.0184 & -0.0149 & 3.2124 & 0.1574\\
		10 & As & 3.1649 & 3.1176 & -0.0065 & 0.4102\\
		11 & As & 3.2514 & 0.0774 & 0.2758 & 0.2669\\
		12 & Bs & 4.0884 & 0.7715 & 1.7469 & 1.6891\\
		13 & Bs & -1.0344 & 1.1210 & 1.3510 & 1.6635\\
		14 & Bs & 3.3374 & 0.1237 & 2.6662 & 0.3175\\
        \hline
        \multicolumn{3}{l}{small-scale: $S=0.025$} &  \multicolumn{3}{c}{Nulls found: $10$}\\
        \hline
		1 & X & 2.9668 & -0.1497 & 0.1298 & 0.1408\\ 
		2 & A & 3.1386 & 3.1344 & -0.0320 & 0.1974\\
		3 & B & 3.4131 & 3.5137 & 2.6837 & 0.7573\\
		4 & B & 3.0098 & 3.0167 & 3.4307 & 0.5873\\
		5 & O & -0.1291 & 3.1253 & 0.2019 & 0.6016\\
		6 & O & 0.0457 & -0.0005 & -0.0104 & 0.0898\\
		7 & O & -0.0006 & 3.1376 & 3.1733 & 0.1556\\
		8 & O & 0.0197 & -0.0166 & 3.2153 & 0.2312\\
		9 & As & 3.1832 & 3.1089 & 0.0113 & 0.5680\\
		10 & Bs & 3.2822 & 0.1041 & 2.8751 & 0.2343\\
	\hline\hline
	\end{tabular}
\end{center}
\end{table}

  \subsubsection{Results of FOTE method}
  Table~\ref{tab:fote} shows the results of the FOTE method testing on null point location and identification. The types, coordinates and minimum distances to spacecraft of these null points are given.

  In the { `medium-scale' tetrahedron configuration}, the spacecraft separation is about $0.12$ {in dimensionless units}. Considering the linear assumption, the null points with the distance (from the tetrahedron centre) less than 1 are reserved. In total, we detected $14$ null points, in which the null points $1$, $2$, $5$, $6$, $8$, $9$, $11$, and $14$ are included in Table~\ref{tab:fote} and thus are real null points. The others are misidentifications, generally with large distances (see null points $4$, $7$, $12$, $13$). This is consistent with the properties of FOTE method. Six real null points (null points $1$, $3$, $7$, $12$, $13$, and $14$ in Table~\ref{tab:nullsexact}) are missed in the test. Among the missed null points, nulls $1$ and $7$ (see Table~\ref{tab:nullsexact}) are located rather far from the spacecraft trajectory, and thus cannot be detected. Null points $12$ and $13$ are close to each other (relative to the spacecraft separation), possibly breaking the linearity of the field in their vicinity. This explains why these null points are not detected by FOTE method.
\begin{figure*}[!hbt]
\begin{center}
    \includegraphics[height=0.7\textheight]{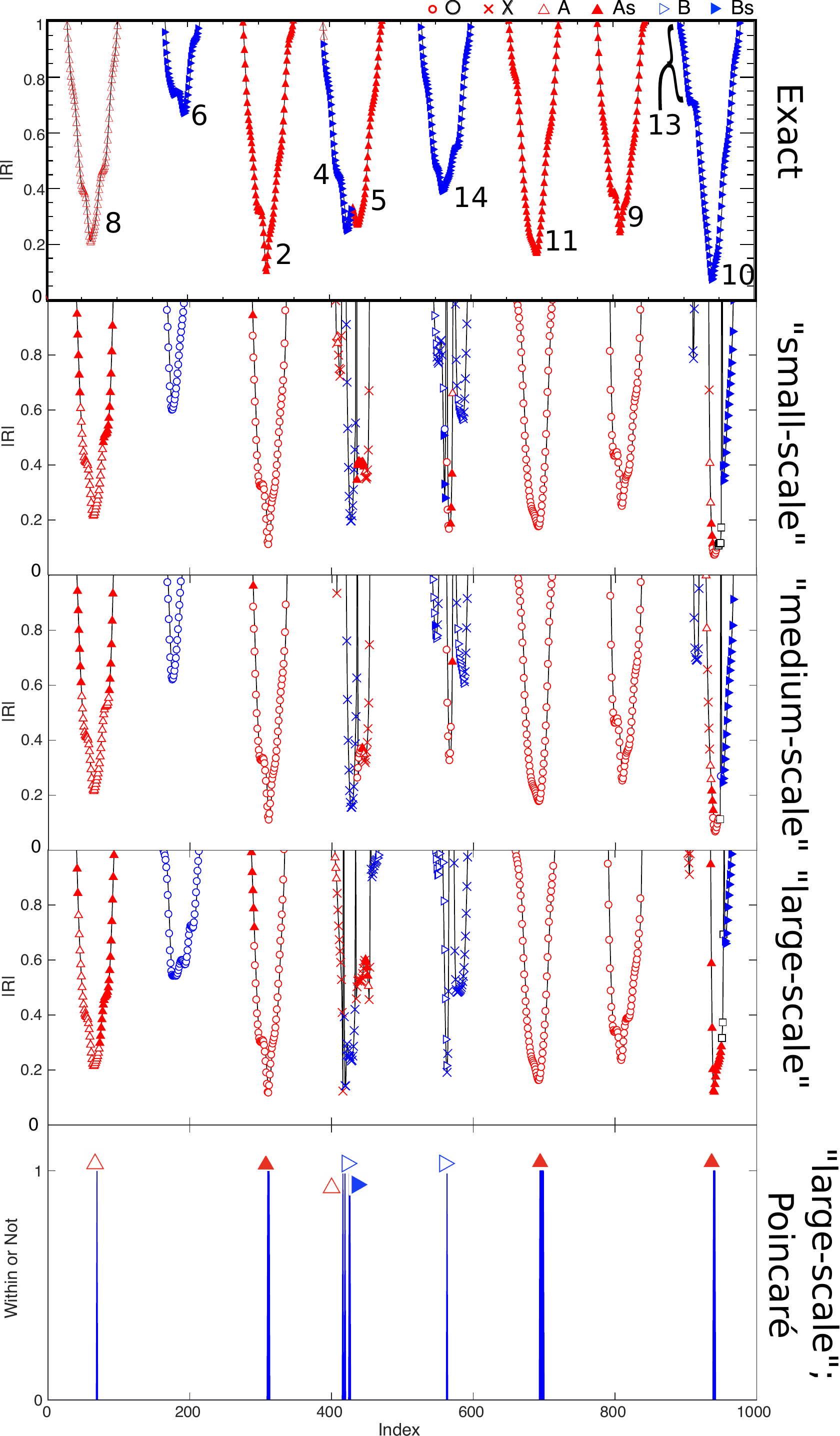}
	\caption{Null points detected based on the simulated spacecraft trajectories as described in Appendix \ref{app:setup}. In the top panel is the exact distance from the centre of the spacecraft tetrahedron to the nearest null {($|R|$)}, its type, and its number based on Table \ref{tab:nullsexact}. The second, third and fourth panels show the results from the FOTE method with a tetrahedron {with  $S=0.025$, $S=0.12$, and $S=0.4$, respectively}. The bottom panel shows the results of applying the Poincar\'e index method {for the large-scale tetrahedron with $S=0.4$}.}
	\label{fig:trajectorytest}
\end{center}
\end{figure*}

  In the {`small-scale' tetrahedron configuration}, the spacecraft separation is about $0.025$. The null points with the distance less than $0.5$ are reserved. In total, we detected 10 null points, in which the null points $1$, $2$, $6$, $7$, $8$, $10$ are included in Table~\ref{tab:fote} and thus are real null points. We note that the null points which are successfully identified by FOTE method are always with the distance less than $0.25$, while the misidentifications are with the distances larger than $0.5$. This means we could conveniently improve the credibility of FOTE method by decreasing the threshold distance.

  The null point detections for the two different tetrahedron sizes  are shown graphically in Fig.~\ref{fig:trajectorytest}, where the exact answer is also plotted in the top panel. In the fourth panel of the Figure we also show the result of applying FOTE to an even larger tetrahedron, for comparison with the Poincar\'e index method (see below).

  One conclusion (which is expected because of the linear-field assumption in the method) is that the FOTE method performs well when the null point is not too close to any other nulls (say, on the scale of the tetrahedron), such as nulls $8$ and $9$. While in the places where two nulls are relatively close together, such as nulls $4$ and $5$, the null type detection is rather erratic.
  
   {There is no clear trend regarding the accuracy of null detections with tetrahedron size. However, it is clear that very large and very small tetrahedron sizes are both bad: when the tetrahedron is very large the field can be far from linear, with many nulls in the local region. On the other hand if it is very small, the field gradients not well sampled, and nulls can be missed because they pass far from the spacecraft on the length-scale of the tetrahedron \citep[and it is known for FOTE from previous experience that for reliable results we should exclude null detections more than a few times the spacecraft separation, e.g.][]{Fu:etal:2020}. The optimal balance, therefore, is to have a spacecraft separation of order  the null separation, but this is not known a priori. In the absence of such knowledge, the optimal size of the tetrahedron can be motivated by some known physical length scales, such as the ion inertial length.}

  \subsubsection{Results of Poincar\'e index method}
  With the {`small-scale' and `medium-scale' tetrahedron configurations} described above, the nulls never pass exactly through the spacecraft tetrahedron. Therefore, to test the Poincar\'e index method we have created trajectories with a `large-scale' tetrahedron (see the Appendix), to ensure the possibility of (true) positive results. The results of applying the Poincar\'e index method to this data set are shown in the bottom panel of Figure~\ref{fig:trajectorytest}. We observe that all nulls that happen to be enclosed by the artificial spacecraft constellation, namely the nulls 8, 2, and 11, have been correctly identified. We have found that, in accordance with \citet{Greene:1992}, the Secant method provides a bad estimate of the enclosed null location, often outside the tetrahedron. Therefore, the best practice is to provide the centre of the tetrahedron as the location of the null point.

  \subsubsection{Comparison between FOTE and Poincar\'e index methods}
  As expected, the FOTE method is able to detect the nulls when they are some distance away from the tetrahedron. Moreover, FOTE detects a null feature in all cases where the null point comes within a distance of 1 from the tetrahedron centre, for all tetrahedron sizes tested. The accuracy of the distance and null type assessment tends to degrade for larger tetrahedron sizes, as expected. The Poincar\'e index method also detects all nulls that could be expected (those that pass through the tetrahedron). The two methods for assessing the type of the null perform similarly, with success rate around 50\%. When multiple nulls are located close together (e.g.~nulls 4 and 5), or too far from the spacecraft (in case of FOTE), both methods detect the presence of nulls, but show noisy results in detection/distance/typing.

  \subsubsection{\bf Tetrahedron trajectory considerations}
  {The results discussed above are based upon data measured along trajectories that traverse a circular path in the $xy$-plane of our domain. Clearly these trajectories do not mimic the behaviour of spacecraft constellations such as Cluster or MMS, which for example in the magnetotail move only slowly as magnetic structures are convected backwards and forwards past them. However, we do not expect the nature of these trajectories to influence the results. The shape of the trajectories is chosen to bring them close to as many of the null points in the domain as possible, in order to test the field reconstruction around each of those nulls, and thus make our analysis more robust. The null point identification is not affected locally by the shape of the trajectory (since only the data at a single time -- or two adjacent times for SOTE -- is used), but rather by the separation of the `spacecraft'. In order to mimic the effects of small-scale fluctuations in the fields and noise, we added the small-scale fluctuations to the trajectories in Equations (\ref{eq:noise1}--\ref{eq:noise3}).
  }


\section{Conclusion}
  This work intends to help researchers who want to analyse null points (stagnation points) of divergence-free vector fields in their simulations or observations. There are two situations that are commonly encountered in practice: (i) numerical simulations on hexahedral or tetrahedral meshes, and (ii) data from tetrahedra of spacecraft (MMS and Cluster). Each of these cases was assessed independently using the same test magnetic field. This is the first time that such methods have been tested and compared against a `ground truth' situation where null numbers, positions, and types were known to arbitrary precision based on an analytical expression for the magnetic field. The main results of our study relevant to `8-point methods' used for rectangular meshes from numerical simulations are the following.

\begin{itemize}
  \item When the field is moderately resolved (1,2, or fewer grid points between nulls), both the Poincar\'e index and trilinear methods give errors, but the trilinear method has no false positives (PI method has 1), fewer false negatives (2 vs 4), and the performance on null type is the same (one incorrect A vs B identification each). This suggests that the trilinear method is more robust when the field is quite `rough' on the grid. 

  \item When the field is well resolved, ($>$2 grid points between nulls) both methods identify all nulls and their types correctly. Since the trilinear method finds the nulls to sub-grid resolution it gets the positions more accurately. The trilinear method does not inherently detect spiral nulls, but could do so by finding Jacobian of the trilinear fit and calculating eigenvalues. The PI method could include an extra step where the fit is made (e.g. trilinear) to get sub-grid resolution of null position.

  \item Both methods can be efficiently implemented to run in less than 1 second on an $80^3$ grid for the present test field, and show no substantial difference in scaling with resolution. 
\end{itemize}

  Concerning 4-point methods typically applied to spacecraft data, {we have considered three different sizes for the spacecraft tetrahedron, the smallest two of which can be considered as `Cluster-scale' and `MMS-scale' on the basis of a physically-motivated dimensionsalisation of the field -- see Section \ref{sec:dataset}.} We conclude that FOTE performs well in finding the nulls when they are not close together -- roughly speaking, when the null separation is larger than the null-spacecraft distance. (The main discrepancies in Figure 5 are the use of X and O for nulls that are close to 2D). On the other hand, for null pairs that are close together the detection method fails (interestingly it still detects a null, but the inferred type jumps around a lot). Probably this could be used to indicate multiple adjacent nulls. 

  The practical advice is to use the FOTE method for locating the nulls in the fields measured by probes or spacecraft. In the numerical simulations on the rectilinear grids the trilinear method gives more accurate null location. On the meshes one should use a variation of the PI on either hexahedral or tetrahedral cells. The null location produced by the latter should be taken in the middle of the cell as the secant method of the location estimation could produce unreliable results.


\section{Discussion}


  We propose that the fields defined in the appendix and used here could be used to test/benchmark future null finders. For example a method based on an expansion in spherical harmonics used by \citep{He:2008,li2019}. The original intention of this method is to reconstruct magnetotail magnetic structure around magnetic null observed by local satellite \citep{He:2008, Guo:2016a}. 
  Based on satellite measurements, it reconstructs the magnetic field by taking advantage of a fitting function approach. To match the 12 observed magnetic field components, 10 fitting parameters are presented in 10 spherical harmonic functions, and the other two are in the Harris current sheet model \citep{Harris:1962NC}. Thus, by fitting the simultaneous magnetic field vectors, one can reconstruct the local magnetic field. The calculations in \cite{He:2008} confirmed the existence of a magnetic field null in reconnection event, and present a magnetic structure around a 3-D null in the magnetotail. For convenience, we provide the theoretical formulation of this method in Appendix~\ref{app:spherical}.

  In order to apply this method, four point measurements are required in the data cube. Any four points that form a tetrahedron in the data box introduced in Section~\ref{sec:background} can be used for reconstruction experiments. For example, in a $80^3$ size data box, one can first choose a $2^3$ data box, and separate it as five independent tetrahedrons. Then five reconstructions can be done based on the tetrahedrons. The following is to check for magnetic nulls in these reconstructed magnetic structures. The advantage is that it can be reconstructed to get multiple nulls, while other methods can't judge the existence of multiple zeros in the area surrounded by multiple satellites. Multiple tetrahedrons can be randomly selected for reconstruction, and the results obtained together with all data points can be compared. However, it would be less efficient if it is used as an ergodic calculation similar to the Poincar\'e index. Also the results of the magnetic nulls need to check the reconstructed magnetic structures manually, which need to be further improved in the future. Once the automated procedure for inspecting and classifying the reconstruction results is developed, this method can be tested against the proposed challenge.


\begin{acknowledgements}
     This work has resulted from ISSI/ISSI-BJ team activity 416 ``Magnetic Topology Effects on Energy Dissipation in Turbulent Plasma".
\end{acknowledgements}

\begin{appendix}
\section{Details of the test magnetic field}\label{app:setup}
The magnetic field model used in this paper is

\begin{eqnarray}
{\bf B}& = (-2\sin(2y)+\sin z){\bf e}_x+(2\sin x+\sin z){\bf e}_y+(\sin x+\sin y){\bf e}_z \nonumber\\
&+0.04(-2\sin(2y)+2\sin x\cos z+\sin(y)+0.1195){\bf e}_x \nonumber\\
&+0.04(2\sin (x-z)+\sin (x+z) + \sqrt{30}/7){\bf e}_y \nonumber\\
&+ 0.04(-2\cos x \sin z+\sin(y)+2\sin(2y)- 0.1378162){\bf e}_z \nonumber\\
&+ \sum_{i=1}^2 \nabla \times\left( a_i k_i \exp\left( -\frac{(x-X_i)^2+(y-Y_i)^2}{a_i^2}-\frac{(z-Z_i)^2}{l_i^2} \right){\bf e}_z \right)\nonumber\\
&+ \sum_{i=3}^5 \nabla \times\left( a_i k_i \exp\left( -\frac{(x-X_i)^2+(z-Z_i)^2}{a_i^2}-\frac{(y-Y_i)^2}{l_i^2} \right){\bf e}_y \right)\nonumber\\
&+ \sum_{i=6}^7 \nabla \times\left( a_i k_i \exp\left( -\frac{(y-Y_i)^2+(z-Z_i)^2}{a_i^2}-\frac{(x-X_i)^2}{l_i^2} \right){\bf e}_x \right) \label{eq:bdef}
\end{eqnarray}
where the values of $a_i$, $k_i$, $l_i$, $X_i$, $Y_i$ and $Z_i$ are given in Table \ref{tab:bparams}.

\begin{table}[!htb]
\centering
\caption{Parameter values for the magnetic field in Equation~(\ref{eq:bdef}).}
\label{tab:bparams}
\begin{tabular}{ccccccc}
$i$ & $a_i$ & $k_i$ & $l_i$ & $X_i$ & $Y_i$ & $Z_i$\\\hline
1 & $-0.2$ & 50 & 0.2 & 2.8455 & 0.0267 & $-0.147$ \\
2 & $-0.3$ & 4 & 0.2 & $\pi+0.191$ & 0.0117 & $\pi$\\
3 & 1 & 0.8 & 1 & 0.1 & $\pi-0.037$ & 0.2\\
4 & $-0.22$ & 1 & 0.9 & $\pi+0.0771$ & $\pi+0.087$ & $\pi$\\
5 & $-0.22$ & 2 & 0.5 & $\pi/2$ & $\pi/2$ & $\pi/2$\\
6 & 1 & 1 & 1 & 0.073 & 0.0198 & $-0.07$\\
7 & $-0.22$ & 2 & 0.5 & $\pi/2$ & $\pi/2$ & $\pi/2$
\end{tabular}
\end{table}

The simulated spacecraft trajectories are constructed as follows. First, a trajectory for the tetrahedron centre is defined. A parametric representation of this curve is given by
\begin{eqnarray}
{\bf r}(s) &=& \left(  \frac{\pi}{\sqrt{2}}\cos(s)+\frac{\pi}{2}+3f_1(s)+0.1\right){\bf e}_x  \nonumber\\
&&+ \left( \frac{\pi}{\sqrt{2}}\sin(s) +\frac{\pi}{2}+3f_2(s) \right){\bf e}_y  \nonumber\\
&&+ \left(  (1-0.025s)\frac{\pi}{2}\tanh(8s)+\frac{\pi}{2}+3f_3(s) \right){\bf e}_z
\end{eqnarray}
where 
\begin{eqnarray}
f_1(s)&=&0.01\,\sin \left( 20\,s+23 \right) + 0.004\,\sin \left( 23\,s+17
 \right) \nonumber\\&& + 0.011\,\sin \left( 13\,s+5 \right) + 0.007\,\sin \left( 37
\,s+13 \right)\label{eq:noise1}\\
f_2(s)&= &0.01\,\sin \left( 19\,s+23 \right) + 0.005\,\sin \left( 25\,s+17
 \right) \nonumber\\&& + 0.009\,\sin \left( 17\,s+5 \right) + 0.013\,\sin \left( 33
\,s+13 \right)\\
f_3(s)&=&0.007\,\sin \left( 22\,s+23 \right) + 0.006\,\sin \left( 24\,s+17
 \right) \nonumber\\&&+ 0.01\,\sin \left( 13\,s+5 \right) + 0.003\,\sin \left( 39\,
s+13 \right) \label{eq:noise3}
\end{eqnarray} 
Next, four constant vectors are added to this this expression to determine four neighbouring trajectories for the four spacecraft:
\begin{eqnarray*}
{\bf V}_1 &=& S\sqrt{3}(0.28881225{\bf e}_x+0.40784216{\bf e}_y+0.28911173{\bf e}_z),\\
{\bf V}_2 &=& S\sqrt{3}(-0.07797703{\bf e}_x-0.45912701{\bf e}_y+0.34125548{\bf e}_z),\\
{\bf V}_3 &=& S\sqrt{3}(-0.51329368{\bf e}_x+0.20729668{\bf e}_y-0.16398482{\bf e}_z),\\
{\bf V}_4 &=& S\sqrt{3}(0.30245845{\bf e}_x-0.15601183{\bf e}_y-0.46638238{\bf e}_z).
\end{eqnarray*}
These vectors lie at the corners of a regular tetrahedron, with each point lying a {\bf distance $S$} from the centre of the tetrahedron {\bf (which has side-length, or spacecraft separation, $S\sqrt{8/3}$)}. Here we consider {\bf three} different tetrahedron sizes, with {\bf $S=0.025$, $S=0.12$ and $S=0.4$, which we refer to as `small-scale', `medium-scale' and `large-scale', respectively}.

The null points within the domain together with the eigenvalues of the associated Jacobian matrix are given in Table \ref{tab:nullsexact}.
\begin{table*}[!hbt]
\begin{center}
	\caption{Null points in the domain and the associated eigenvalues - exact values.}
	\label{tab:nullsexact}
	\begin{tabular}{cc|ccc|ccc}
    	\hline\hline
    	
        {Index} & Type & x & y & z & $\lambda_1$ & $\lambda_2$ & $\lambda_3$ \\\hline
 $1$&$  B_s $&$ 0.2952 $&$ -0.2505 $&$ -1.4803 $&$ -0.7018 $&$  0.3509+2.3321j $&$ 0.3509-2.3321j$\\
 $2$&$ A_s  $& $0.04581 $&$ -0.0005391 $&$ -0.01064 $&$ 0.8959 $&$-0.4479+3.3971j $&$ -0.4479-3.3971j  $\\
 $3$&$  A $&$2.8657 $&$-0.1878$&$-0.5961$ &$  4.8311$&$ -0.8373$&$ -3.9938 $\\
 $4$&$  B_s $&$2.9737$&$  -0.1500$&$ 0.2246$&$ -8.0909 $&$  4.0454+1.4497j $&$ 4.0454-1.4497j$\\
 $5$&$  A_s $&$3.2286 $&$  0.06090$&$ 0.2457 $&$  2.9331 $&$ -1.4666+1.3798j $&$ -1.4666-1.3798j$\\
$6$&$  B_s $&$ -0.2124 $&$ 3.3212 $&$ 0.4444 $&$ -0.6310$&$ 0.3155+3.1129j $&$ 0.3155-3.1129j $\\
 $7$&$  B_s $&$ -0.2728$&$ 0.2386 $&$ 1.2466 $&$ -0.8733 $&$  0.4367+2.1948j $&$ 0.4367-2.1948j$\\
$8$&$  A $&$3.1391$&$  3.1344$&$  -0.03510 $&$  2.9710 $&$ -0.8713 $&$ -2.0998$\\
$9$&$  A_s $&$0.01951 $&$ -0.01637 $&$ 3.2123 $&$  0.2230$&$ -0.1115+3.1491j$&$ -0.1115-3.1491j $\\
  $10$&$  B_s $&$ 3.3225$&$  0.1453$&$ 2.9898 $&$ -0.3269 $&$  0.1635+13.3803j $&$ 0.1635-13.3803j$\\
  $11$&$  A_s $&$ -0.0017705 $&$ 3.1360$&$ 3.1707 $&$  0.7427 $&$ -0.3713+2.9089j $&$ -0.3713-2.9089j $\\
 $12$&$  A_s $&$ 3.3119$&$  0.1539$&$ 3.2509 $&$  0.4726 $&$ -0.2363+17.6743j $&$ -0.2363-17.6743j$\\
$13$&$  B_s $&$2.9915$&$  -0.09684$&$  3.5047 $&$ -2.6941 $&$  1.3470+0.9660j $&$ 1.3470-0.9660j$\\
$14$ &$B_s$&$3.0203$&$  3.0182$&$  3.3935 $&$-4.00896$ & $2.0449+0.8020j$& $2.0449-0.8020j$\\
 
	\hline\hline
	\end{tabular}
\end{center}
\end{table*}


\section{Spherical expansion method formulation}
\label{app:spherical}

The fitting model is designed based on a total of 12 functions, including a constant background field, a function taken from the Harris current sheet model \citep{Harris:1962NC}, and 10 spherical harmonic functions. For the convenience of describing the potential field, the spherical harmonic functions are adopted as a part of the fitting model. Such fitting can be expressed as
\begin{equation}
    \begin{pmatrix}
        B_\gamma \\
        B_\theta \\
        B_\phi
    \end{pmatrix} = 
    \begin{pmatrix}
        \widetilde{B_r} \\
        \widetilde{B_\theta} \\
        \widetilde{B_\phi}
    \end{pmatrix} + 
    T_{xyz \rightarrow \gamma\theta\phi} \cdot 
    \begin{pmatrix}
        B_0 \tanh{\frac{z-z_0}{L_z}} + B_1 \\
        \widetilde{B_\theta} \\
        \widetilde{B_\phi}
    \end{pmatrix},
\label{eq:sperical:01}    
\end{equation}
where $\left( B_\gamma,\, B_\theta,\, B_\phi \right)$ are the three spherical coordinate system magnetic field components at a spatial position $\left( \gamma,\,\theta,\,\phi\right)$. The first term on the right-hand side describes a potential field from the spherical harmonic series shown below. The transform matrix $T_{xyz \rightarrow \gamma\theta\phi}$. converts vector field from a common spacecraft Cartesian coordinate system to a spherical coordinate system. The x-direction background magnetic field together with the magnetic field in a Harris current model is shown in this equation. Expression for $\widetilde{B_r}$, $\widetilde{B_\theta}$, $\widetilde{B_\phi}$ is shown as
\begin{align*}
      \widetilde{B_r} &=  \sum_{n}\sum_{m}{-\left(n+1\right)\left(\frac{R_e}{r}\right)^{n+2}\cdot\left(q_n^m\cos{\left(m\varphi\right)}+h_n^m\sin{\left(m\varphi\right)}\right)}\\
      &\qquad\qquad \cdot\ P_n^m\left(\cos{\theta}\right) \\
      \widetilde{B_\theta}&=  \frac{R_e}{r}\sum_{n}\sum_{m}{\left(\frac{R_e}{r}\right)^{n+1}\cdot\left(q_n^m\cos{\left(m\varphi\right)}+h_n^m\sin{\left(m\varphi\right)}\right)}\cdot\left(-\sin{\theta}\right)\\
      &\qquad\qquad \cdot\frac{\partial}{\partial\theta}\left(P_n^m\left(\cos{\theta}\right)\right) \\
       \widetilde{B_\phi}&= \frac{R_e}{r\sin{\theta}}\sum_{n}\sum_{m}{\left(\frac{R_e}{r}\right)^{n+1}\cdot\left(q_n^m\left(-m\right)\cdot\sin{\left(m\varphi\right)}+h_n^m\cdot m\cdot\cos{\left(m\varphi\right)}\right)}\\
      &\qquad\qquad \cdot P_n^m\left(\cos{\theta}\right),
\end{align*}
where $P_n^m$ is the associated Legendre function with degree n and order $m$ of $\left[n, m\right] = {[1,1], [2,1], [{2,2}], [3,1], [3,2]}$, and $q_n^m$ and $h_n^m$ are the coefficients in the spherical harmonic series.

\section{Null-finder validation tools}
  Data files and source codes for testing null-finders are available online at https://bitbucket.org/volshevsky/magneticnullchallenge.
\end{appendix}

\bibliographystyle{aa} 

\begin{thebibliography}{41}
\expandafter\ifx\csname natexlab\endcsname\relax\def\natexlab#1{#1}\fi

\bibitem[{Burch {et~al.}(2016)Burch, Moore, Torbert, \& Giles}]{Burch:2016}
Burch, J.~L., Moore, T.~E., Torbert, R.~B., \& Giles, B.~L. 2016, Space Science
  Reviews, 199, 5

\bibitem[{Chen {et~al.}(2017)Chen, Fu, Liu, Cao, Wang, Dunlop, Chen, \&
  Peng}]{Chen:etal:2017}
Chen, X.~H., Fu, H.~S., Liu, C.~M., {et~al.} 2017, The Astrophysical Journal,
  852, 17

\bibitem[{Chen {et~al.}(2019)Chen, Fu, Wang, Liu, \& Xu}]{Chen:etal:2019}
Chen, Z.~Z., Fu, H.~S., Wang, Z., Liu, C.~M., \& Xu, Y. 2019, Geophysical
  Research Letters, 46, 10209

\bibitem[{Close {et~al.}(2003)Close, Parnell, MacKay, \& Priest}]{close2003}
Close, R.~M., Parnell, C.~E., MacKay, D., \& Priest, E.~R. 2003, Solar Phys.,
  212, 251

\bibitem[{{Deng} {et~al.}(2009){Deng}, {Zhou}, {Li}, {Baumjohann}, {Andre},
  {Cornilleau}, {Santol{\'{\i}}k}, {Pontin}, {Reme}, {Lucek}, {Fazakerley},
  {Decreau}, {Daly}, {Nakamura}, {Tang}, {Hu}, {Pang}, {B{\"u}chner}, {Zhao},
  {Vaivads}, {Pickett}, {Ng}, {Lin}, {Fu}, {Yuan}, {Su}, \& {Wang}}]{deng2009}
{Deng}, X.~H., {Zhou}, M., {Li}, S.~Y., {et~al.} 2009, J.~Geophys.~Res., 114,
  A07216

\bibitem[{{Eriksson} {et~al.}(2015){Eriksson}, {Vaivads}, {Khotyaintsev},
  {Khotyayintsev}, \& {Andr{\'e}}}]{Eriksson:etal:2015}
{Eriksson}, E., {Vaivads}, A., {Khotyaintsev}, Y.~V., {Khotyayintsev}, V.~M.,
  \& {Andr{\'e}}, M. 2015, \grl, 42, 6883

\bibitem[{Escoubet {et~al.}(2001)Escoubet, Fehringer, \&
  Goldstein}]{Escoubet:2001}
Escoubet, C.~P., Fehringer, M., \& Goldstein, M. 2001, Annales Geophysicae, 19,
  1197

\bibitem[{Fu {et~al.}(2019)Fu, Cao, Cao, Wang, Vaivads, Khotyaintsev, Burch, \&
  Huang}]{Fu:etal:2018}
Fu, H.~S., Cao, J.~B., Cao, D., {et~al.} 2019, Geophysical Research Letters,
  46, 48

\bibitem[{Fu {et~al.}(2017)Fu, Vaivads, Khotyaintsev, André, Cao, Olshevsky,
  Eastwood, \& Retinò}]{Fu:etal:2016}
Fu, H.~S., Vaivads, A., Khotyaintsev, Y.~V., {et~al.} 2017, Geophysical
  Research Letters, 44, 37

\bibitem[{{Fu} {et~al.}(2015){Fu}, {Vaivads}, {Khotyaintsev}, {Olshevsky},
  {Andr{\'e}}, {Cao}, {Huang}, {Retin{\`o}}, \& {Lapenta}}]{Fu:etal:2015JGR}
{Fu}, H.~S., {Vaivads}, A., {Khotyaintsev}, Y.~V., {et~al.} 2015, Journal of
  Geophysical Research (Space Physics), 120, 3758

\bibitem[{Fu {et~al.}(2020)Fu, Wang, Zong, Chen, He, Vaivads, \&
  Olshevsky}]{Fu:etal:2020}
Fu, H.~S., Wang, Z., Zong, Q., {et~al.} 2020, Methods for Finding Magnetic
  Nulls and Reconstructing Field Topology (American Geophysical Union (AGU)),
  153--172

\bibitem[{Fukao {et~al.}(1975)Fukao, Ugai, \& Tsuda}]{Fukao:1975}
Fukao, S., Ugai, M., \& Tsuda, T. 1975, Rep. Ion. Sp. Res. Japan, 29, 133

\bibitem[{{Greene}(1992)}]{Greene:1992}
{Greene}, J.~M. 1992, Journal of Computational Physics, 98, 194

\bibitem[{Guo {et~al.}(2016)Guo, Pu, Chen, Fu, Xie, Wang, Dunlop, Bogdanova,
  Yao, Xiao, He, \& Fazakerley}]{Guo:2016a}
Guo, R., Pu, Z., Chen, L.-J., {et~al.} 2016, Physics of Plasmas, 23, 052901

\bibitem[{{Harris}(1962)}]{Harris:1962NC}
{Harris}, E.~G. 1962, Il Nuovo Cimento, 23, 115

\bibitem[{{Haynes} \& {Parnell}(2007)}]{haynesparnell2007}
{Haynes}, A.~L. \& {Parnell}, C.~E. 2007, Physics of Plasmas, 14, 082107

\bibitem[{He {et~al.}(2008)He, Tu, Tian, Xiao, Wang, Pu, Ma, Dunlop, Zhao,
  Zhou, Wang, Fu, Liu, Zong, Glassmeier, Reme, Dandouras, \&
  Escoubet}]{He:2008}
He, J.-S., Tu, C.-Y., Tian, H., {et~al.} 2008, Journal of Geophysical Research:
  Space Physics, 113
  [\eprint{https://agupubs.onlinelibrary.wiley.com/doi/pdf/10.1029/2007JA012609}]

\bibitem[{Hornig \& Schindler(1996)}]{Hornig:1996}
Hornig, G. \& Schindler, K. 1996, Phys. Plasmas, 3, 781

\bibitem[{{Khurana} {et~al.}(1996){Khurana}, {Kepko}, {Kivelson}, \&
  {Elphic}}]{Khurana:etal:1996IEEE}
{Khurana}, K., {Kepko}, E., {Kivelson}, M., \& {Elphic}, R. 1996, Magnetics,
  IEEE Transactions on, 32, 5193

\bibitem[{{Kumar} {et~al.}(2019){Kumar}, {Karpen}, {Antiochos}, {Wyper},
  {DeVore}, \& {DeForest}}]{kumar2019}
{Kumar}, P., {Karpen}, J.~T., {Antiochos}, S.~K., {et~al.} 2019, \apj, 873, 93

\bibitem[{{Li}(2019)}]{li2019}
{Li}, S. 2019, International Journal of Geosciences, 10, 967

\bibitem[{Liu {et~al.}(2019)Liu, Fu, Olshevsky, Pontin, Liu, Wang, Chen, Dai,
  \& Retino}]{Liu:2019:SOTE}
Liu, Y.~Y., Fu, H.~S., Olshevsky, V., {et~al.} 2019, The Astrophysical Journal
  Supplement Series, 244, 31

\bibitem[{{Longcope} \& {Parnell}(2009)}]{longcope2009}
{Longcope}, D.~W. \& {Parnell}, C.~E. 2009, Solar Phys., 254, 51

\bibitem[{Masson {et~al.}(2009)Masson, Pariat, Aulanier, \&
  Schrijver}]{masson2009}
Masson, S., Pariat, E., Aulanier, G., \& Schrijver, C.~J. 2009, \apj, 700, 559

\bibitem[{{Olshevsky} {et~al.}(2016){Olshevsky}, {Deca}, {Divin}, {Peng},
  {Markidis}, {Innocenti}, {Cazzola}, \& {Lapenta}}]{Olshevsky:2016ApJ}
{Olshevsky}, V., {Deca}, J., {Divin}, A., {et~al.} 2016, \apj, 819, 52

\bibitem[{{Olshevsky} {et~al.}(2015){Olshevsky}, {Divin}, {Eriksson},
  {Markidis}, \& {Lapenta}}]{Olshevsky:etal:2015ApJ}
{Olshevsky}, V., {Divin}, A., {Eriksson}, E., {Markidis}, S., \& {Lapenta}, G.
  2015, \apj, 807, 155

\bibitem[{{Olshevsky} {et~al.}(2018){Olshevsky}, {Servidio}, {Pucci},
  {Primavera}, \& {Lapenta}}]{Olshevsky:etal:2018ApJ}
{Olshevsky}, V., {Servidio}, S., {Pucci}, F., {Primavera}, L., \& {Lapenta}, G.
  2018, \apj, 860, 11

\bibitem[{Parnell {et~al.}(1996)Parnell, Smith, Neukirch, \&
  Priest}]{Parnell:1996}
Parnell, C.~E., Smith, J.~M., Neukirch, T., \& Priest, E.~R. 1996,
  Phys.~Plasmas, 3, 759

\bibitem[{Politano {et~al.}(1995)Politano, Pouquet, \& Sulem}]{Politano:1995}
Politano, H., Pouquet, A., \& Sulem, P.~L. 1995, Physics of Plasmas, 2, 2931

\bibitem[{Pontin(2012)}]{pontin2012}
Pontin, D.~I. 2012, Phil.~Trans.~R.~Soc.~A, 370, 3169

\bibitem[{Priest(2014)}]{priest2014}
Priest, E.~R. 2014, Magnetohydrodynamics of the Sun (Cambridge University
  Press, Cambridge.)

\bibitem[{{Pucci} {et~al.}(2017){Pucci}, {Servidio}, {Sorriso-Valvo},
  {Olshevsky}, {Matthaeus}, {Malara}, {Goldman}, {Newman}, \&
  {Lapenta}}]{Pucci:etal:2017ApJ}
{Pucci}, F., {Servidio}, S., {Sorriso-Valvo}, L., {et~al.} 2017, \apj, 841, 60

\bibitem[{{Retin{\`o}} {et~al.}(2007){Retin{\`o}}, {Sundkvist}, {Vaivads},
  {Mozer}, {Andr{\'e}}, \& {Owen}}]{Retino:etal:2007}
{Retin{\`o}}, A., {Sundkvist}, D., {Vaivads}, A., {et~al.} 2007, Nature
  Physics, 3, 236

\bibitem[{Schrijver \& Title(2002)}]{schrijver2002}
Schrijver, C.~J. \& Title, A.~M. 2002, Solar Phys., 207, 223

\bibitem[{{van Oosterom} \& {Strackee}(1983)}]{VanOosterom:Strackee:1983IEEE}
{van Oosterom}, A. \& {Strackee}, J. 1983, Biomedical Engineering, IEEE
  Transactions on, 30, 125

\bibitem[{Wang {et~al.}(2020)Wang, Fu, Olshevsky, Liu, Liu, \&
  Chen}]{Wang:etal:2020}
Wang, Z., Fu, H.~S., Olshevsky, V., {et~al.} 2020, The Astrophysical Journal
  Supplement Series, 249, 10

\bibitem[{{Wendel} \& {Adrian}(2013)}]{wendel2013}
{Wendel}, D.~E. \& {Adrian}, M.~L. 2013, Journal of Geophysical Research (Space
  Physics), 118, 1571

\bibitem[{Williams(2018)}]{williams2018}
Williams, B.~M. 2018, PhD thesis, University of St Andrews

\bibitem[{{Wyper} \& {Pontin}(2014)}]{Wyper:2014a}
{Wyper}, P.~F. \& {Pontin}, D.~I. 2014, Phys.~Plasmas, 21, 082114

\bibitem[{{Xiao} {et~al.}(2006){Xiao}, {Wang}, {Pu}, {Zhao}, {Wang}, {Ma},
  {Fu}, {Kivelson}, {Liu}, {Zong}, {Glassmeier}, {Balogh}, {Korth}, {Reme}, \&
  {Escoubet}}]{Xiao:etal:2006NatPh}
{Xiao}, C.~J., {Wang}, X.~G., {Pu}, Z.~Y., {et~al.} 2006, Nature Physics, 2,
  478

\bibitem[{{Yang} {et~al.}(2015){Yang}, {Guo}, \& {Ding}}]{yang2015}
{Yang}, K., {Guo}, Y., \& {Ding}, M.~D. 2015, Astrophys.~J., 806, 171

\end{thebibliography}

\end{document}